\documentclass
[aps,preprint,showkeys,a4paper,tightenlines,superscriptaddress]{revtex4}%
\usepackage{eurosym}
\usepackage{amsfonts}
\usepackage{amsmath}
\usepackage{amssymb}
\usepackage{graphicx}
\usepackage{subfig}
\usepackage{caption}%
\providecommand{\U}[1]{\protect\rule{.1in}{.1in}}
\providecommand{\U}[1]{\protect\rule{.1in}{.1in}}

\begin{document}
\title{The nonlinear dynamic behavior of a Rubber-Layer Roller Bearing (RLRB) for
vibration isolation}
\author{N. Menga}
\email{[Corresponding author. ]Email: nicola.menga@poliba.it, phone number: +39 080 5962746}
\affiliation{Department of Mechanics, Mathematics and Management, Politecnico of Bari, V.le
Japigia, 182, 70126, Bari, Italy}
\affiliation{Imperial College London, Department of Mechanical Engineering, Exhibition
Road, London SW7 2AZ}
\author{F. Bottiglione}
\affiliation{Department of Mechanics, Mathematics and Management, Politecnico of Bari, V.le
Japigia, 182, 70126, Bari, Italy}
\author{G. Carbone}
\affiliation{Department of Mechanics, Mathematics and Management, Politecnico of Bari, V.le
Japigia, 182, 70126, Bari, Italy}
\affiliation{Imperial College London, Department of Mechanical Engineering, Exhibition
Road, London SW7 2AZ}
\affiliation{CNR - Institute for Photonics and Nanotechnologies U.O.S. Bari, Physics
Department \textquotedblright M. Merlin\textquotedblright, via Amendola 173,
70126 Bari, Italy}
\keywords{Nonlinear dynamics, viscoelastic damping, seismic isolation, base isolation}
\begin{abstract}
In this paper, we study the dynamic behavior of a Rubber-Layer Roller Bearing
(RLRB) interposed between a spring-mass elemental superstructure and a
vibrating base. Thanks to the viscoelastic rolling contact between the rigid
rollers and the rubber layers, the RLRB is able to provide a nonlinear damping
behavior. The effect of the RLRB geometric and material parameters is
investigated under periodic base excitation, showing that both periodic and
aperiodic responses can be achieved. Specifically, since the viscoelastic
damping is non-monotonic (bell shaped), there exist system dynamic conditions
involving the decreasing portion of the damping curve in which a strongly
nonlinear behavior is experienced. In the second part of the paper, we
investigate the effectiveness of the nonlinear device in terms of seismic
isolation. Focusing on the mean shock of the Central Italy 2016 earthquake, we
opportunely tune the material and geometrical RLRB parameters, showing that a
significant reduction of both the peak and root-mean-square value of the
inertial force acting on the superstructure is achieved, compared to the best
performance of a linear base isolation system.

\end{abstract}
\maketitle

\section{Introduction}

Absorbing and controlling the vibration of mechanical systems and structure is
a very demanding task. Due to the always increasing demand from mechanical,
aeronautical and civil engineering, the last decades have seen a proliferation
of applications of\ nonlinear systems to vibration control, thanks to their
intrinsic ability (i) to efficiently react to external forcing in a much wider
range of frequency compared to linear systems, (ii) to modify their behavior
according to the excitation amplitude.

In this view, one of the most common vibration absorption strategy relies on
the adoption of nonlinear energy sinks (NES), whose mechanism depends on the
specific field of application. To this regard, in Ref. \cite{Starosvetsky2009}
the effect of the nonlinear (quadratic) damping behavior given by an hydraulic
damper equipped with several on/off valves is studied, showing that the
removal of unwanted periodic regimes can be achieved by means of opportunely
tuned damping characteristics. Similar studies were then extended to the case
of vibro-impact NES \cite{Gendelman2012,Gendelman2015}, showing that chaotic
dynamic regimes are easy to occur, thus promoting these systems for energy
harvesting applications. Similarly, focusing on the case of travelling loads
on elastic beams (e.g. railway tracks under moving trains), an extensive study
on nonlinear tuned mass dampers has been performed in Refs.
\cite{Samani2009,Samani2012,Samani2013}, showing that stiffness nonlinearity
poorly affects the overall dynamic response, whereas nonlinear damping may
lead to great vibration reduction. Moreover, the case of a moving mass-damper
for transmission cables is investigated in Ref. \cite{Bukhari2018}, where it
is shown that significantly higher energy dissipation can be achieved compared
to the case of fixed dampers.

Among the application fields of vibration control, seismic engineering is one
of the fastest growing sector, as the ability to ensure high reliability for
primary structures and machineries (e.g., power plants, hospitals, schools,
etc.) has strong social, political, and economic implications
\cite{Myslimaj2003}. Therefore, several passive systems have been developed to
deal with this task, mostly relying on nonlinear stiffness behavior. Indeed,
both inter-story frictional dissipators \cite{DelaCruz2007} for high-rise
buildings (where the source of nonlinearity are the frictional interactions),
and bi-component sacrificial supports \cite{Foti2013} able to provide a
piecewise linear foundation stiffness have shown general vibration absorption
and reduced structure response.

Base isolation systems are also well established methods to control the
superstructure dynamic response, as indeed reviewed in Ref. \cite{Harvey2016}.
Among them, very promising solutions relies rolling isolation systems (RIS),
where the rolling of rigid balls on concave counterface provides a nonlinear
gravitational stiffness, and external viscous dashpot provides linear damping.
In Refs. \cite{Harvey2015,Casey2018}, it is shown that such devices are
peculiarly suited for heavy low-rise structures, presenting significantly
enhanced isolation performances. On the same path, the idea to combine passive
base isolation with active structural control is investigated in Ref.
\cite{Miyamoto2018}, where several control logics are explored in order to
minimize a specific performance index based on absolute acceleration, and
inter-story drift and velocity.

On the other hand, less effort has been paid to study in details the dynamic
character of such isolation systems. Indeed, only a few works highlighted
that, under specific conditions, RIS may show chaotic behavior
\cite{Harvey2013,Harvey2013b} (mostly due to the variable curvature of the
rolling counterfaces), thus resulting in significant sensitivity to initial
conditions and, in turn, less engineering predictability of the overall
isolation behavior.

Furthermore, in order to provide specific nonlinear stiffness and damping,
most of the RISs for base isolation systems rely on separated mechanical
components \cite{Lin1995} (e.g. nonlinear spring, hydraulic dampers with
valves, etc.), which need to be arranged in specific, and usually complex,
configurations. This entails high installation and maintenance costs, as well
as reduced reliability.

\begin{figure}[ptbh]
\centering\includegraphics[width=0.5\textwidth]{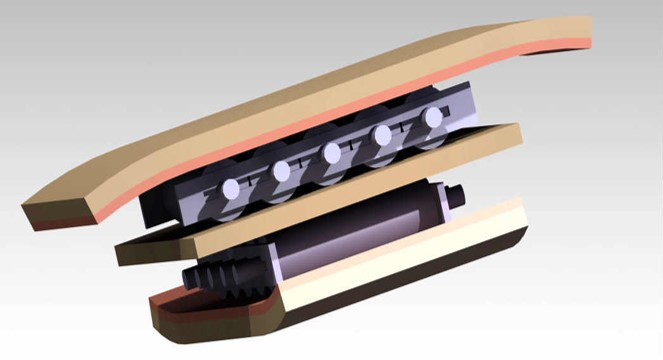}\caption{A
representation of a bi-axial cylindrical RLRB: two perpendicular rows of
equispaced steel rollers are interposed between steel plates coated with
rubber layers (from Ref. \cite{menga2017}).}%
\label{fig1}%
\end{figure}

To overcome these limitations, Rubber-Layer Roller Bearings (RLRB) have been
proposed so far \cite{Foti1996,Foti1996b,Muhr1997,MuhrConf,Guerreiro2007},
where the rolling balls counterparts were opportunely coated with highly
viscoelastic rubber, thus providing both significant damping without
additional devices. Similar studies were later widened to the case of rolling
rods in Refs. \cite{Foti2013b,patentFoti2015,Foti2019}. Such devices are
specifically suited to provide a significant base-isolation degree in the case
of horizontal excitations. Notably, as reported in Refs.
\cite{Mazza2016,Mazza2018}, in the case of near-fault earthquake the seismic
excitation may involve long-period horizontal pulses in the fault-normal
velocity signals and high values of the ratio between the peak of the vertical
and horizontal ground accelerations. Under these conditions, RLRB systems may
suffer uplifts in the vertical direction, thus resulting in vanishing damping.
However, compared with similar friction-based isolation devices (e.g. RIS and
friction pendulum), due to the rubber coating, RLRBs present lower vertical
stiffness (i.e. larger vertical isolation), which may allow, in turn, to
tolerate higher vertical peak acceleration without experiencing complete uplifts.

Although pioneeristic, the studies performed on RLRBs do not provide a
detailed insight into the viscoelastic bulk dissipation mechanism, and in turn
into the damping behavior of such systems, which is instead addressed by means
of phenomenological models. To this regard, in a recent paper \cite{menga2017}%
, we deeply investigated the damping behavior of an innovative RLRB based on
rolling cylinders (see fig. \ref{fig1}), which, showing overall lower contact
pressures (see also Ref. \cite{Foti2013b}), provides higher rubber reliability
compared to sphere-based RLRB. In the framework of linear viscoelasticity, we
accurately defined the damping curve associated to viscoelastic bulk
dissipation, assuming steady rolling conditions.

In this paper, we try to widen the investigation of cylindrical RLRBs by
studying the dynamic behavior of the exemplar case of a single-story
superstructure base isolated by means of a RLRB. The paper is organized in two
sections. The first one is devoted to the viscoelastic contact mechanics
formulation, based on Boundary Element Method (BEM) with specific viscoelastic
Green's function taking into account for the system materials and geometry,
together with the dynamic model of the system, where the two degree of freedom
equations of motion are derived. In the second section, we present our main
results: firstly we focus on the system dynamics under periodic base
excitation, highlighting the effect of the RLRB geometrical and material
parameters; then, a detailed optimization of the RLRB parameters is performed
considering a real earthquake base excitation. To stress the effect of the
specific RLRB behavior nonlinearity on the system response, the results are
compared with the case of an equivalent linear base isolation device.

\section{Formulation}

\subsection{The RLRB viscoelastic behavior}

\begin{figure}[ptbh]
\centering\includegraphics[width=0.5\textwidth]{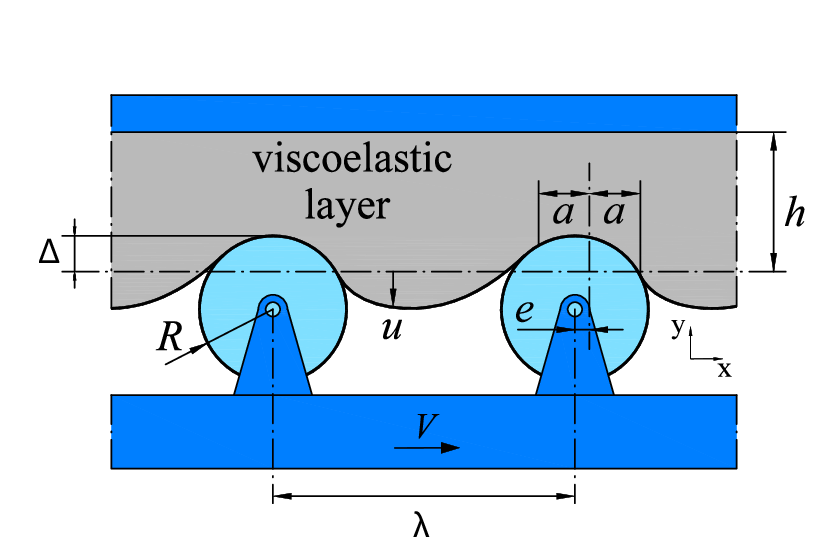}\caption{The
geometrical scheme of the periodic rolling contact under investigation. In
particular, $\Delta$ is the contact penetration between the cylinders and the
deformed surface mean plane, and $u$ is the layer local displacement. Due to
the material viscoelasticity, the contact area mean line is shifted by a
quantity $e$ with respect to the cylinders axis.}%
\label{fig2}%
\end{figure}

The damping behavior of the base isolation systems depends on the viscoelastic
contact behavior of the RLRB. In fig. \ref{fig2}, we show a portion of the
periodic contact between the RLRB equispaced rigid rollers (of radius $R$) and
the viscoelastic layer stuck onto a rigid plate. The relative motion between
the top and the bottom plate leads to cyclic deformations of the viscoelastic
rubber coating, entailing bulk dissipation which gives rise to a reaction
force, opposing the relative motion. In what follows, we assume frictionless
contact between the rigid cylinders and the viscoelastic layer of thickness
$h$. Since our study is developed within the framework of linear
viscoelasticity, we neglect any large deformation effect. Referring to fig.
\ref{fig2}, for a given value of the velocity $V$, following Ref.
\cite{menga2016visco}, the mean shear stress acting on the upper body can be
easily calculated as%
\begin{equation}
f_{m}\left(  V\right)  =-\frac{1}{\lambda}\int_{\Omega}p\left(  x\right)
u^{\prime}\left(  x\right)  dx \label{fmean}%
\end{equation}
where $\lambda$ is the periodic distance between the cylinders, $p\left(
x\right)  $ is the contact pressure distribution, and $u^{\prime}\left(
x\right)  $ the first derivative of the displacement field of the viscoelastic
layers within the contact domain $\Omega=\left[  -a,a\right]  $, being $a$ the
semi-width of the contact area (see Fig. \ref{fig2}).

Eq. (\ref{fmean}) clearly shows that the overall damping force depends on the
contact pressure distribution, which is unknown. Following Ref.
\cite{menga2017}, by exploiting the symmetry of the system, we focus our study
on half of the device, as indeed shown in fig. \ref{fig2}. Furthermore, it can
be demonstrated that $f_{m}\left(  V\right)  $ is an odd function of $V$, as
the mean tangential shear stress always opposes the relative motion between
the RLRB upper and lower parts.

Following the procedure delineated in Refs.\cite{carb-mang-2008,menga2017}, by
relying on the Green's function approach, the displacement and the contact
pressure fields can be related by means of a specific Green's function, which,
in the case of steady sliding, parametrically depends on $V$. Thus,
\begin{equation}
u\left(  x\right)  =-\int_{\Omega}\Theta_{V}\left(  x-\xi\right)  p\left(
\xi\right)  d\xi. \label{1}%
\end{equation}
The kernel $\Theta_{V}\left(  x\right)  $ is the viscoelastic Green's function
for steady sliding contacts, which has been already calculated in the case of
periodic contacts with layers of finite thickness in Refs.
\cite{menga2016visco,menga2018visco}. We report herein the main relations,
assuming linear viscoelastic material with a single relaxation time $\tau$%
\begin{equation}
\Theta_{V}\left(  x\right)  =\frac{1}{E_{\infty}}G\left(  x\right)  +\frac
{1}{E_{1}}\int_{0^{+}}^{+\infty}G\left(  x+V\tau\rho\right)  \exp\left(
-\rho\right)  d\rho
\end{equation}
where $1/E_{1}=1/E_{0}-1/E_{\infty}$, being $E_{0}$ and $E_{\infty}$
respectively the zero-frequency and high frequency elastic moduli of the
material. The elastic-like Green's function is related to the specific
geometry under investigation, and takes the form
\begin{equation}
G\left(  x\right)  =\frac{2\left(  1-\nu^{2}\right)  }{\pi}\log\left[
2\left\vert \sin\left(  \frac{kx}{2}\right)  \right\vert \right]
+\frac{2\left(  1-\nu^{2}\right)  }{\pi}\sum_{m=1}^{\infty}A_{m}\left(
kh\right)  \frac{\cos\left(  mkx\right)  }{m} \label{4}%
\end{equation}
with $k=2\pi/\lambda$ and%

\begin{equation}
A_{m}\left(  kh\right)  =\frac{2hkm-\left(  3-4\nu\right)  \sinh\left(
2hkm\right)  }{5+2\left(  hkm\right)  ^{2}-4\nu\left(  3-2\nu\right)  +\left(
3-4\nu\right)  \cosh\left(  2hkm\right)  }+1
\end{equation}
where $\nu$ is the material Poisson's ratio.

Fig. \ref{fig2} shows that, due the viscoelastic delay in the material
response, the contact area exhibit a\ certain degree of eccentricity $e$ with
respect to the mean line of the cylinder cross-section. Moreover, within the
contact strip $\Omega$, \ the layer displacement must copy the rigid cylinder
shape, i.e. $u\left(  x\right)  =\Delta-\Lambda\left[  1-r\left(  x+e\right)
\right]  $, where $r\left(  x\right)  =R\sin\left[  \cos^{-1}\left(
x/R\right)  \right]  $ is the profile of the upper half-cylinder, and
$\Lambda=R-\lambda^{-1}\int_{\lambda}r\left(  x\right)  dx$. Under these
conditions, Eq. (\ref{1}) represents a Fredholm equation of the first kind
which is solved for the unknown contact pressure distribution by exploiting
the numerical scheme already discussed in Refs.
\cite{menga2016,menga2019coated,menga2019coupling} for adhesiveless contacts.

\subsection{The system dynamics}

RLRB devices are usually adopted to achieve a certain degree of dynamic base
isolation \cite{Harvey2016} between the ground (e.g. seismic) motion and
several superstructures, such as buildings, machinery, etc. Since in this
study we are interested in highlighting the dynamic behavior peculiarities of
the RLRB system, we focus on a very simple superstructure: an elastic pillar
(with bending stiffness $k_{2}$) supporting an inertial mass $m_{2}$.

\begin{figure}[ptbh]
\centering\includegraphics[width=0.7\textwidth]{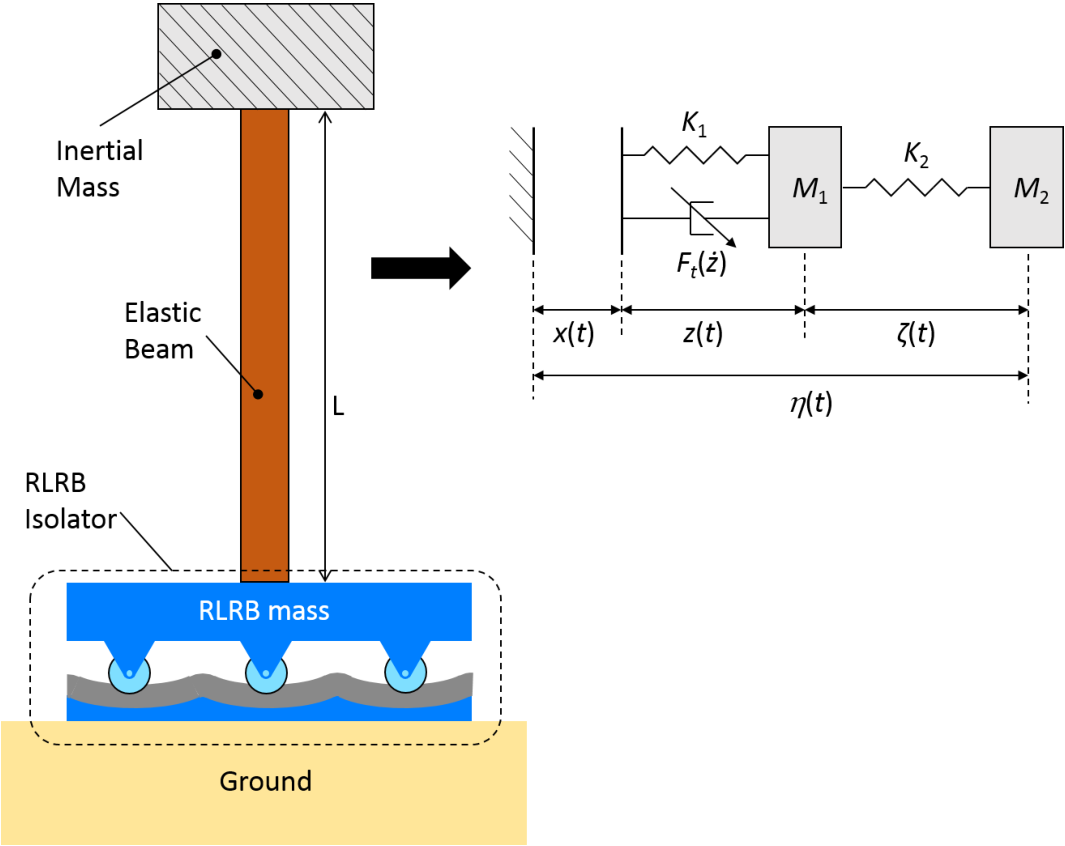}\caption{A sketch
of the base-isolated physical system. The heavy mass $m_{2}$ is connected to
the RLRB system by means of an elastic beam of length $L$ and bending
stiffness $K_{2}$. The RLRB rolling path is concave, thus resulting in a
linearized recentering stiffness $K_{1}$. On the right: a lumped element
scheme where $x$ is the ground absolute displacement, $z$ and $\zeta$ are the
relative displacement between the ground and the top RLRB plate, $\,$and RLRB
and inertial mass, respectively.}%
\label{fig3}%
\end{figure}

A functional scheme of the resulting system is shown in figure \ref{fig3},
together with a lumped element picture of the system. Specifically, we
consider the case of a concave RLRB (with radius of curvature $R_{b}\gg R$),
whose gravitational re-centering effect is taken into account by means of a
linearized gravitational spring with stiffness
\begin{equation}
k_{1}=g\frac{m_{1}+m_{2}}{R_{b}} \label{k1}%
\end{equation}
where $g$ is the gravitational acceleration.

The dissipation of the viscoelastic rolling contact occurring between the RLRB
rigid cylinders and rubber layer leads to a damping tangential force opposing
the relative motion between the superstructure and the ground. Such a force
can be calculated as
\[
F_{d}\left(  \dot{z}\right)  =-N\lambda b\left\vert f_{m}\left(  \dot
{z}\right)  \right\vert \frac{\dot{z}}{\left\vert \dot{z}\right\vert }%
\]
where $N$ is the number of rigid cylinders of the RLRB device, and $b$ is the
transverse width of the systems.

The equations of motion of the system of fig. \ref{fig3} are
\begin{equation}
\left\{
\begin{array}
[c]{c}%
m_{1}\left(  \ddot{x}+\ddot{z}\right)  +k_{1}z-F_{d}\left(  \dot{z}\right)
-k_{2}\zeta=0\\
m_{2}\left(  \ddot{x}+\ddot{z}+\ddot{\zeta}\right)  +k_{2}\zeta=0
\end{array}
\right.  \label{eqsys}%
\end{equation}
where $x\left(  t\right)  $ is the ground vibration. We also define
\begin{equation}
\eta(t)=x(t)+z(t)+\zeta(t)
\end{equation}
as the absolute displacement of the inertial mass.

Eqs. (\ref{eqsys}) represent a set of nonlinear second order ODE, which have
been integrated numerically by relying on a fixed time-step method based on
fourth order \textit{Runge-Kutta} algorithm \cite{Demidovich}. To avoid
numerical instabilities, a sensibility study has been performed on the effect
of the time-step value on the integration result.

Interestingly, the equilibrium along the vertical direction of the physical
system shown in fig. \ref{fig3} allows us to calculate the contact mean
pressure acting on the rigid cylinders-rubber layer interface as%
\begin{equation}
p_{m}=\frac{1}{\lambda}\int_{\Omega}p\left(  x\right)  dx=g\frac{m_{1}+m_{2}%
}{N\lambda b} \label{pm}%
\end{equation}

Notably, due to the oscillatory shape of the base excitation $x(t)$ typical of
seismic and vibrational phenomena, the RLRB undergoes to a reciprocating
motion. In this case, the viscoelastic contact between the rubber layer and
the rigid cylinders belongs to the class of reciprocating rolling contacts,
usually requiring sophisticated theoretical treatments to address the specific
frictional and contact behavior. However, in Ref. \cite{reciprocating2016}, it
has been shown that simplified unidirectional steady motion analysis may still
provide good qualitative and quantitative predictions, depending on the actual
operating conditions. In particular, once defined the linear size $2a$ of the
contact area between the cylinders and the rubber, the stroke $s$ and the
period $T$ of the reciprocating motion, the viscoelastic contact behavior
closely resembles the one observed in steady sliding at constant velocity
provided that $a\ll s$ and $\tau\ll T$. Since in our analysis, the latter
conditions are met, we will here approximate the reciprocating viscoelastic
response with the equivalent unidirectional steady one.

\section{Results}

\subsection{Viscoelastic contact behavior}

In this section, we present the main results in term of contact conditions
experienced at the interface between the rubber layer and the rigid cylinders.
Specifically, we consider the case of a single relaxation time incompressible
viscoelastic material (i.e. $\nu=0.5$), whose high and zero frequency elastic
moduli are $E_{\infty}=150$ MPa and $E_{0}=50$ MPa, respectively.

\begin{figure}[ptbh]
\centering\includegraphics[width=0.55\textwidth]{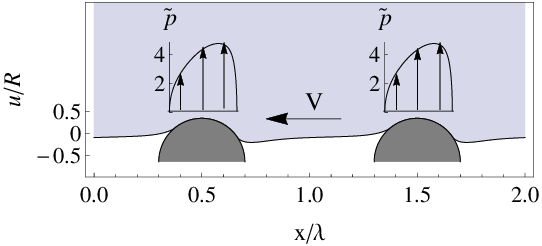}\caption{The
deformed contact configuration under rolling at steady velocity.}%
\label{fig4}%
\end{figure}

Fig. \ref{fig4} shows a typical shape of the deformed layer in steady rolling
contact over cylindrical indenters (i.e. the rigid rollers). A certain degree
of eccentricity of the contact area with respect to the rigid cylinders
meanline is experienced due to the delay in the material response (i.e. the
energy dissipation occurring in the bulk viscoelastic material), which, in
turn, also gives rise to asymmetric contact pressure distributions. Although
dealing with namely frictionless contact, under these conditions, following
Eq. (\ref{fmean}), the contact force presents a tangential component (the
so-called "viscoelastic friction") which opposes to the relative motion
between the cylinders and the rubber layer. Focusing on our physical system
(see fig. \ref{fig3}) such a force represent the damping force opposing the
motion between the superstructure and the ground.

\begin{figure}[ptbh]
\centering\subfloat[\label{fig5a}]{\includegraphics[height=5.2cm]{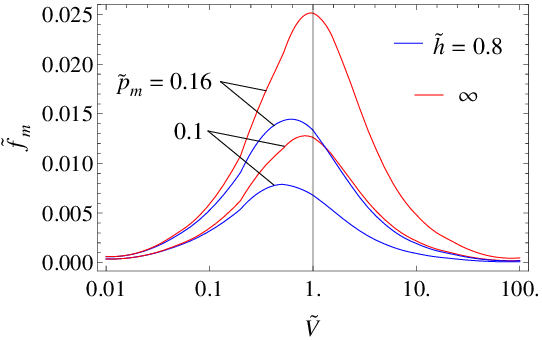}}\hfill
\subfloat[\label{fig5b}]{\includegraphics[height=5.2cm]{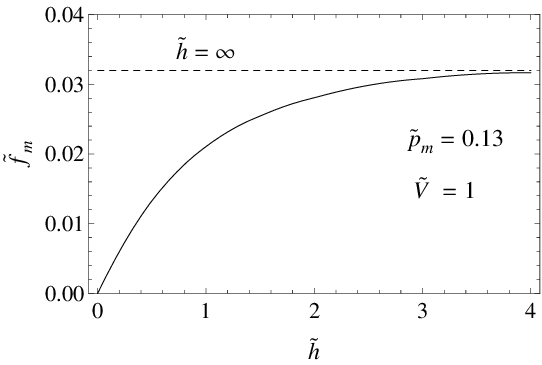}}\caption{The
dimensionless mean shear stress $\tilde{f}_{m}$ as a function of (a) the
dimensionless relative velocity $\tilde{V}$, and (b) the dimensionless
thickness $\tilde{h}$. Results refer to (a) $\tilde{R}=0.3$, and (b)
$\tilde{R}=0.1$}%
\label{fig5}%
\end{figure}

Figs. \ref{fig5} show the frictional viscoelastic behavior of the contact.
Specifically, from fig. \ref{fig5a}, showing the dimensionless friction mean
shear stress $\tilde{f}_{m}=2\left(  1-\nu^{2}\right)  f_{m}/E_{0}$ as a
function of the dimensionless velocity $\tilde{V}=V\tau k$ for different
values of the dimensionless contact mean pressure $\tilde{p}_{m}=2\left(
1-\nu^{2}\right)  p_{m}/E_{0}$, we observe that the viscoelastic friction
follows the well-known bell shaped curve, as at very high and very low
excitation frequency it behaves as an elastic material, with vanishing bulk
dissipation. On the contrary, in the range of intermediate frequency, the
viscous dissipation plays a key role, and the frictional force arises. This is
because the largest viscoelastic energy dissipation, and hence friction,
occurs when $Im[E(\omega)]/|E(\omega)|$ is maximized, i.e. at values of
$\omega\tau\approx1$ (see Ref. \cite{menga2016visco,menga2018visco}), where
$\omega$ is the excitation frequency, and
\begin{equation}
E\left(  \omega\right)  =E_{0}+E_{1}\frac{\mathrm{i}\omega\tau}{1+\mathrm{i}%
\omega\tau}\label{complex_modulus}%
\end{equation}
is the viscoelastic complex modulus, with $E_{1}=E_{\infty}-E_{0}$.

Furthermore, from fig. \ref{fig5a} we observe that the thicker the rubber
layer, the larger the friction value is, as increasing $\tilde{h}$ the amount
of deformed material increases as well, leading to higher bulk dissipation.
This is more clearly shown in Fig. \ref{fig5b}, where $\tilde{f}_{m}$ is
plotted as a function of the dimensionless thickness $\tilde{h}=kh$. Of
course, increasing the rubber layer thickness, the viscoelastic half-plane
behavior (i.e. for $h=\infty$) is asymptotically recovered, whereas, in the
limit of vanishing thickness (i.e. for $h\rightarrow0$), vanishing
viscoelastic friction is achieved.

\subsection{Dynamic behavior with periodic base excitation}

Let us now focus on the dynamic response of the physical system under periodic
base excitation in the form $x\left(  t\right)  =A_{0}\sin\left(  \omega
t\right)  $, where $A_{0}$ and $\omega$ are, respectively, the amplitude and
the frequency of the excitation. The physical system we focus on is typical of
seismic engineering, thus we set $m_{1}=1\times10^{2}$ kg, $m_{2}%
=1\times10^{5}$ kg. We assume a concave shape for the RLRB pathway with width
$b=1$ m, and radius of curvature $R_{b}=3.3$ m, which from Eq. \ref{k1} gives
$k_{1}=3\times10^{5}$ N/m. We also assume $\tilde{h}=0.8$ and $\tilde{R}=0.3$.
Similarly, the elastic pillar is constituted by a commercial HEB 300 steel
beam, with $L=3$ m, whose bending stiffness is $k_{2}=6\times10^{6}$ N/m.
Moving from these values, the modal analysis of the system allows to identify
the two natural frequencies $\omega_{1}=1.69$ rad/s, and $\omega_{2}$ $=251$ rad/s.

\begin{figure}[ptbh]
\centering\includegraphics[width=0.5\textwidth]{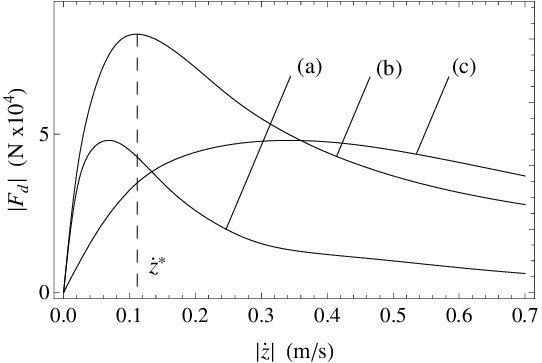}\caption{The
absolute value of the viscoelastic damping force $\left\vert F_{d}\right\vert
$ as a function of the absolute relative velocity $\left\vert \dot
{z}\right\vert $ between the ground and the upper plate of the RLRB device.
(a) $\tau=0.25$ s, $\lambda=0.25$ m; (b) $\tau=0.05$ s, $\lambda=0.05$ m; (c)
$\tau=0.05$ s, $\lambda=0.25$ m. Notably, $\dot{z}^{\ast}$ corresponds to the
peak force (only shown for (b)).}%
\label{fig6}%
\end{figure}

Regarding the viscoelastic nonlinear damping force, coherently with the
dimensionless results presented in fig. \ref{fig5a}, we observe that, given
the values of $\tilde{h}$ and $\tilde{R}$, the final load-velocity curve
depends on both the values of $\tau$ and $\lambda$. The effect of such
parameters on the damping behavior of the RLRB is shown in fig. \ref{fig6}. We
observe that reducing the ratio $\tau/\lambda$ leads to lower slope of the
curve close to the origin. Similarly, increasing $\lambda$ causes a reduction
of the peak force value as, through Eq. (\ref{pm}), it entails a reduction of
the contact mean pressure, thus reducing the overall amount of material
involved in the cyclic deformation, and in turn the energy dissipation.
Notably, $\dot{z}^{\ast}$ defined as the absolute value of the velocity
corresponding to the peak force, depends on the specific parameters as well.

\begin{figure}[ptbh]
\centering\includegraphics[width=0.5\textwidth]{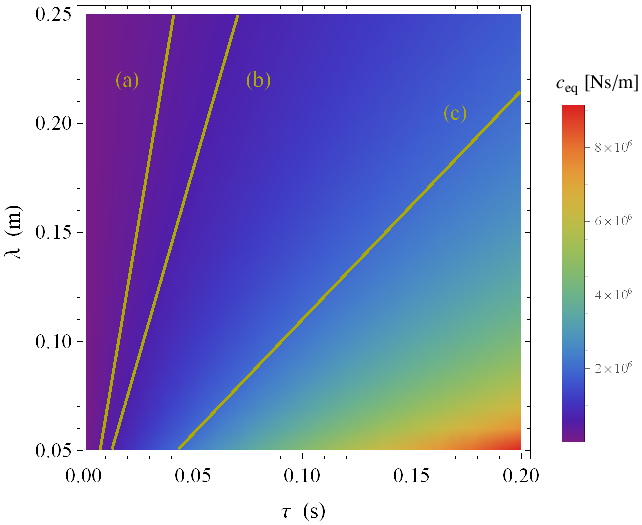}\caption{The
equivalent linear damping coefficient $c_{eq}$ as \ a function of $\tau$ and
$\lambda$. The three lines are for (a) $c_{eq}=3.5\times10^{5}$ Ns/m; (b)
$c_{eq}=6\times10^{5}$ Ns/m; (a) $c_{eq}=2\times10^{6}$ Ns/m; }%
\label{fig7}%
\end{figure}

Furthermore, for any specific damping curve, it is possible to define an
"equivalent" linear viscous damping behavior with damping coefficient%
\begin{equation}
c_{eq}=\left.  -\frac{\mathrm{d}F_{d}}{\mathrm{d}V}\right\vert _{V=0}%
\end{equation}

Fig. \ref{fig7} shows the effect of $\tau$ and $\lambda$ on the value of
$c_{eq}$, in a contour plot. We observe that, according to fig. \ref{fig6},
increasing the viscoelastic relaxation time $\tau\,$, as well as reducing
$\lambda$, the equivalent damping coefficient increases. Further, in the same
figure, three lines have been added referring to specific values of $c_{eq}$,
each one allowing to determine a sets of $\tau$ and $\lambda$ whose equivalent
linear damping behavior is the same. Of course, we expect the system dynamics
to be strongly affected by the values of $\tau$ and $\lambda$, as the strongly
nonlinear viscoelastic damping of the RLRB may lead to completely different
behaviors even in the case of similar linearized equivalent damping coefficients.

\begin{figure}[ptbh]
\centering\includegraphics[width=0.95\textwidth]{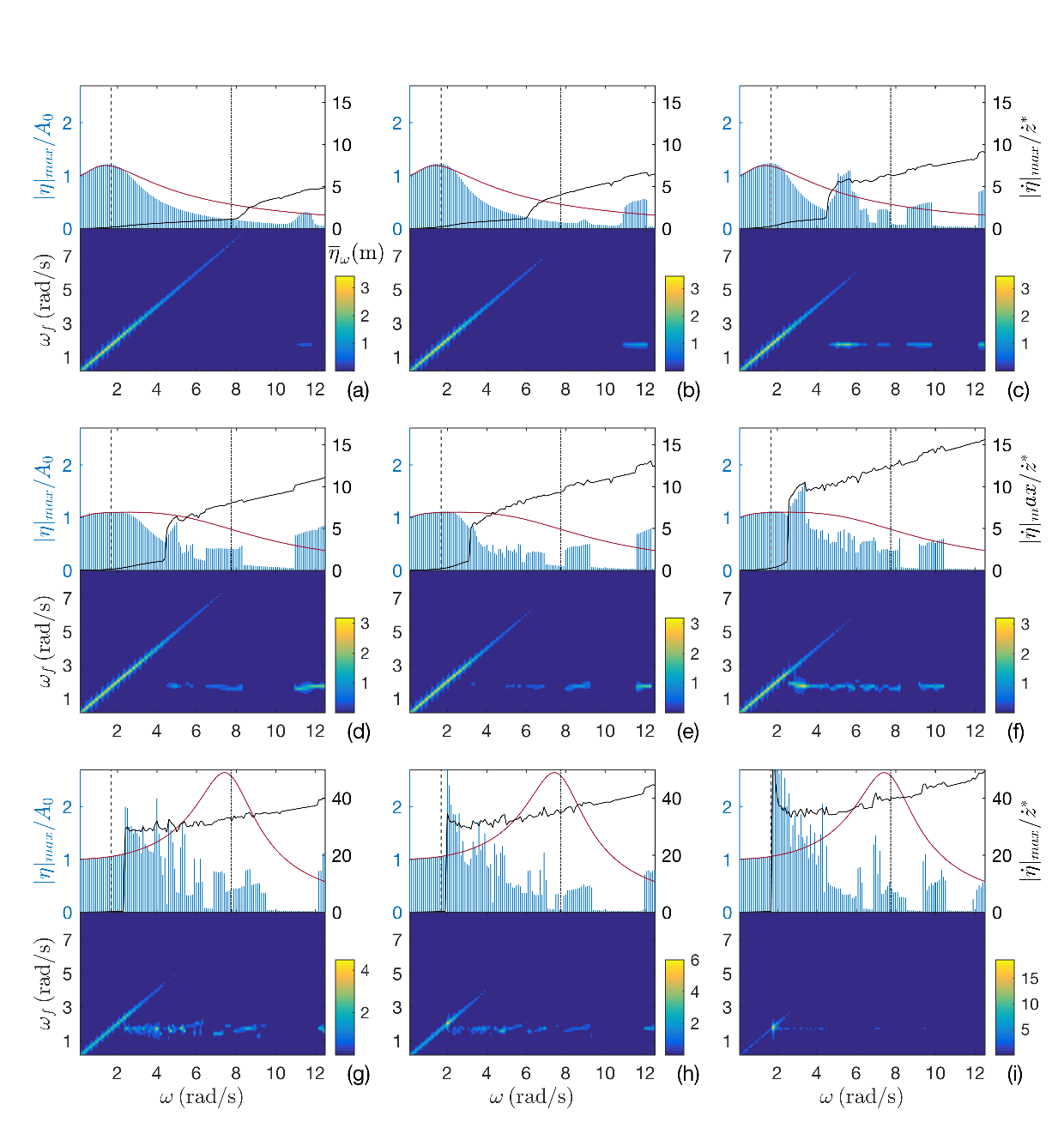}\caption{Left
axis: the dimensionless peak amplitude of the system response $\left\vert
\eta\right\vert _{\max}$ for the non-linear (blue histogram) and equivalent
linearized (red curve) systems as a function of the excitation frequency
$\omega$. Right axis: the dimensionless peak amplitude of the system velocity
response $\left\vert \dot{\eta}\right\vert _{\max}$ (black curve). \ Lower
contour plots, the system response spectrum $\bar{\eta}\left(  \omega
_{f}\right)  $ as a function of the excitation frequency $\omega$.}%
\label{fig8}%
\end{figure}

This is clearly shown in fig. \ref{fig8} where nine sets of values of $\tau$
and $\lambda$ are investigated, as detailed in Table \ref{tab1}. Specifically,
each row of figures refers to the same linearized equivalent damping
coefficient, increasing from the top to the bottom. The dashed vertical lines
represent the system first natural frequency $\omega_{1}$, and the dash-dotted
lines correspond to the natural frequency of an equivalent rigid-damper
system. The figures show, in the lower part, a contour plot of the
steady-state system response spectrum $\bar{\eta}_{\omega}\left(  \omega
_{f}\right)  =\mathcal{F}\left(  \eta\left(  t,\omega\right)  \right)  $ as a
function of the excitation frequency $\omega$ (where $\mathcal{F}$ is the
Fourier transform operator). Similarly, in the upper part we show, on the left
axis, the steady-state maximum amplitude $\left\vert \eta\right\vert _{\max}$
of the system response for the nonlinear (blue histogram) and equivalent
linearized (red curve) systems as a function of the excitation frequency
$\omega$, whereas on the right axis (black curve) the steady-state maximum
amplitude $\left\vert \dot{\eta}\right\vert _{\max}$ of the system velocity
response is shown.

\begin{table}[ptb]
\centering%
\begin{tabular}
[c]{|c|c|c|c|c|}\hline
Case & $\ \ \ \tau$ $\ [s]$ \ \ \ \  & $\ \ \ \lambda$ \ $[m]$ \ \ \  &
$\ c_{eq}$ \ $[Ns/m]$ & $\ \dot{z}^{\ast}$ \ $[m/s]$\\\hline
(a) & $0.00765$ & $0.05$ & $3.5\times10^{5}$ & $0.724$\\
(b) & $0.024$ & $0.15$ & $3.5\times10^{5}$ & $0.534$\\
(c) & $0.041$ & $0.25$ & $3.5\times10^{5}$ & $0.417$\\\hline
(d) & $0.0131$ & $0.05$ & $6\times10^{5}$ & $0.423$\\
(e) & $0.0411$ & $0.15$ & $6\times10^{5}$ & $0.312$\\
(f) & $0.0702$ & $0.25$ & $6\times10^{5}$ & $0.244$\\\hline
(g) & $0.0437$ & $0.05$ & $2\times10^{6}$ & $0.127$\\
(h) & $0.12$ & $0.132$ & $2\times10^{6}$ & $0.097$\\
(i) & $0.2$ & $0.214$ & $2\times10^{6}$ & $0.081$\\\hline
\end{tabular}
\caption{RLRB characteristics of fig. \ref{fig8}}%
\label{tab1}%
\end{table}

Figures \ref{fig8}a, \ref{fig8}b, \ref{fig8}c share the same linearized
behavior (i.e. $\tau/\lambda$ is almost constant) with maximum amplitude of
oscillation close to the system natural frequency $\omega_{1}$. Interestingly,
moving from fig. \ref{fig8}a to \ref{fig8}c the value of $\lambda$ increases
(see the data in Table \ref{tab1}) thus the nonlinear damping force peak value
reduces. This entails that the larger the value of $\lambda$, the smaller the
value of $\omega$ at which the system operating conditions overcome the
damping force peak threshold velocity $\dot{z}^{\ast}$. Indeed, strongly
nonlinear effects usually occurs only at sufficiently large values of $\omega$
where the system response involves $\left\vert \dot{z}\right\vert _{\max}%
>\dot{z}^{\ast}$(i.e. on the decreasing portion of the damping force curve of
fig. \ref{fig6}); whereas, at sufficiently small value of $\omega$, the
nonlinear system vibration closely resembles the one of the linearized system.

Such a peculiar behavior is even more clearly shown by figures \ref{fig8}g,
\ref{fig8}h, \ref{fig8}i, all referring to a set of $\tau$ and $\lambda$
associated to a very high value of linearized equivalent damping coefficient
$c_{eq}$. Under these conditions, since the linear viscous damper behaves
almost rigidly, the linear system behavior (see the red curves) closely
resembles the one of a one degree of freedom harmonic oscillator, of undamped
mass $m_{2}$ and stiffness $k_{2}$ (i.e. maximum amplitude of oscillation
close to the natural frequency $\sqrt{k_{2}/m_{2}}=7.75$ rad/s). The nonlinear
system behaves differently. Indeed, for the specific parameters, the nonlinear
damping force peak value is reached even for low excitation frequency, and
strong nonlinear effects occurs. Notably, from the system response spectrum
shown in the lower contour plots, we observe that the overall response of the
non linear system always involves a harmonic term associated to the external
periodic excitation, whereas the main effect of the nonlinear damping is to
"chaotically" switch on the harmonic term related to the low natural frequency
$\omega_{1}$.

\begin{figure}[ptbh]
\centering\includegraphics[width=0.92\textwidth]{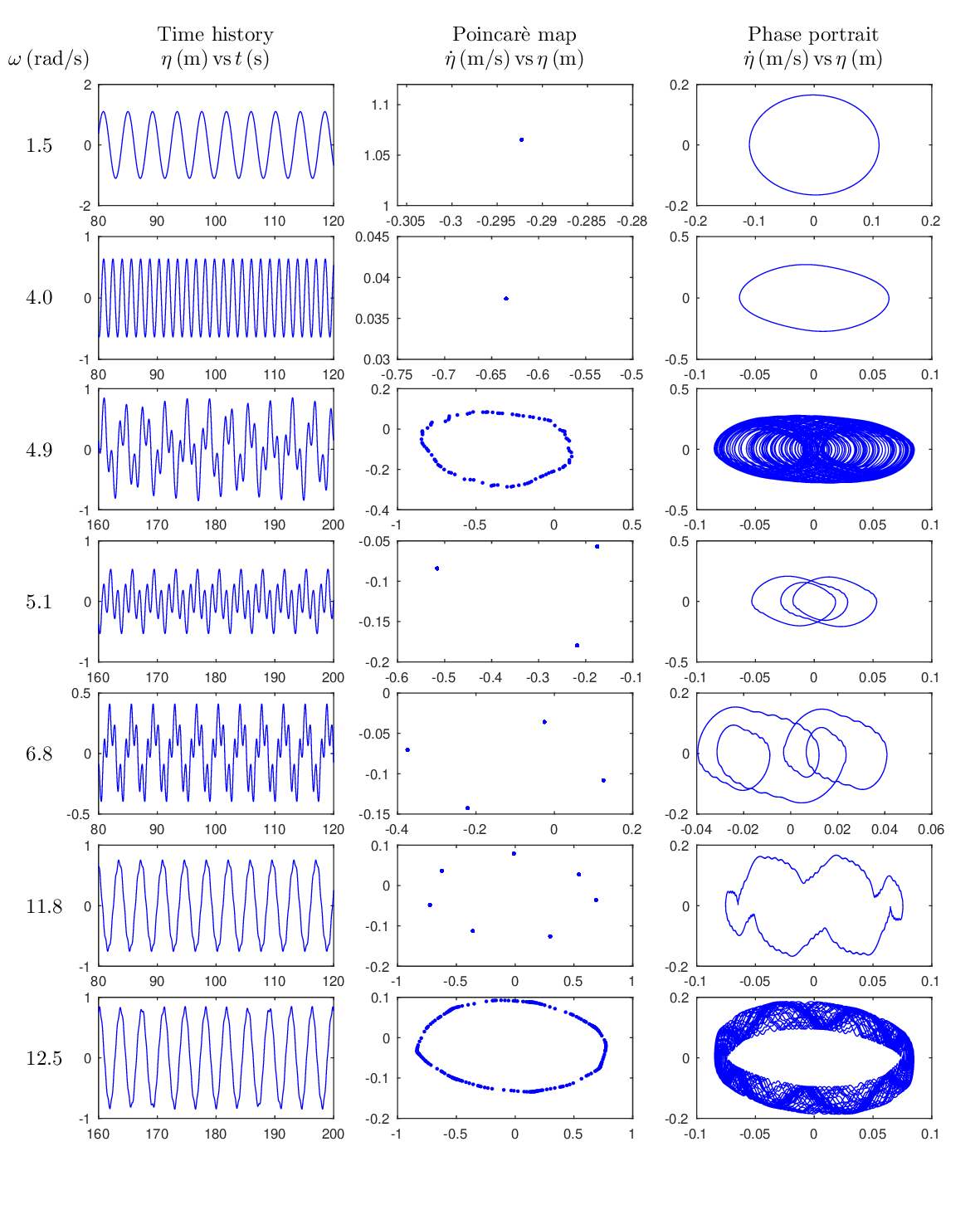}\caption{The
system dynamic behavior for case (d) of Table \ref{tab1}, at varying
excitation frequency $\omega$.}%
\label{fig9}%
\end{figure}

Figure \ref{fig9} illustrates the system behavior of case (d) of Table
\ref{tab1} at different excitation frequencies. Both the system response time
histories, phase portraits and Poincar\'{e} maps refers to the steady-state
conditions. The Poincar\'{e} maps (or recurrence maps) have been achieved by
sampling the system response at intervals equal to the excitation period, with
random phase.

According to fig. \ref{fig8}d, at $\omega=1.5$ rad/s the system behaves
linearly with a periodic response, thus the Poincar\'{e} map is one single
point, and the phase portrait is an ellipse.

Increasing $\omega$ up to $4$ rad/s, nonlinear damping starts to play a
nonvanishing role. The system response is still periodic, but the phase
portrait is now a deformed ellipse.

At $\omega=4.9$ rad/s the system response involves $\left\vert \dot
{z}\right\vert _{\max}>\dot{z}^{\ast}$ (see fig. \ref{fig8}d). Under these
conditions, the vibration spectrum is the sum of two main incommensurable
harmonics: the first excitation harmonic $\omega$, and the system low natural
frequency $\omega_{1}$. As a consequence, the resulting Poincar\'{e} map is a
closed curve, whereas the phase portrait is not, filling a portion of the
phase space.

A slight increase of $\omega$ up to $5.1$ rad/s leads to different results.
This time the ratio of the main frequencies of the system response spectrum is
an integer number, as $\omega/\omega_{1}\approx3$. Since the response is
periodic the phase portrait is a closed curve, and the Poincar\'{e} map shows
the same number of isolated points such as the ratio $\omega/\omega_{1}$. A
similar behavior is also shown at $\omega=6.8$ rad/s and $\omega=11.8$ rad/s,
where respectively $\omega/\omega_{1}\approx4$ and $\omega/\omega_{1}\approx7$

Increasing $\omega$ up to $12.5$ rad/s, the two main components become
incommensurable, the system dynamics is not periodic and a closed curve is
observed in the Poincar\'{e} map associated with a colored region in the phase portrait.

\section{Optimization of the RLRB dynamic behavior}

In the previous section it has been clearly pointed out that the dynamic
behavior of the physical system is strongly affected by the specific damping
behavior of the RLRB base isolation device. Since the latter depends, in turn,
by the specific choice of the physical parameters $\tau$ and $\lambda$, it is
evident that a fine tuning can be performed in order to optimize the overall
behavior of the system with respect to a performance index.

To stress the impact of our conclusions, in what follows, we focus on a real
seismic event, namely the main shock of the Central Italy earthquakes
\cite{seismicdata} occurred on October 30$^{th}$ 2016, with magnitude 6.6
M$_{w}$. In terms of performance indexes, most of the previous studies focuses
on multi-story superstructure \cite{Preumont2008}, in which the main source of
damage is the inter-story drift, leading to critical shear stresses, and
eventually to the structural collapse. In these cases, the most adopted
performance indexes are the relative velocity and displacement of each story
\cite{Ng2007,Yanik2014}. However, since our work is more fundamental, we
define a performance index $\phi$ which encompasses two source of damage for
the structural elements (i.e. the elastic beam of our system): (i) the maximum
inertial load $F_{\max}^{i}$ on the mass $m_{2}$, associated to the structure
instantaneous damage \cite{Sadek1998}; (ii) the root mean square $F_{rms}^{i}$
of the inertial loads history during the shake, associated to the material
hysteresis and fatigue. Specifically, we have that
\begin{equation}
\phi=\frac{1}{2}\left(  \frac{F_{\max}^{i}}{F_{\max,0}^{i}}+\frac{F_{rms}^{i}%
}{F_{rms,0}^{i}}\right)  \label{funzobj}%
\end{equation}
where%
\begin{equation}
F_{\max}^{i}=m_{2}\left\vert \ddot{\eta}\right\vert _{\max}%
\end{equation}
being $\left\vert \ddot{\eta}\right\vert _{\max}$ the absolute acceleration
maximum, and
\begin{equation}
F_{rms}^{i}=m_{2}\sqrt{\int_{t_{1}}^{t_{2}}\frac{\ddot{\eta}\left(  t\right)
^{2}}{t_{2}-t_{1}}dt}%
\end{equation}

The optimization strategy is the following. Firstly, single objective
minimization of $F_{\max}^{i}\left(  \tau,\lambda\right)  $ and $F_{rms}%
^{i}\left(  \tau,\lambda\right)  $ are set independently. The homogenization
terms $F_{\max,0}^{i}$ and $F_{rms,0}^{i}$, in Eq. (\ref{funzobj}), are then
defined as the corresponding values in single objective optimized conditions.
Finally, the minimization of $\phi\left(  \tau,\lambda\right)  $ is performed.

\begin{figure}[ptbh]
\centering\subfloat[\label{fig10a}]{\includegraphics[height=5.2cm]{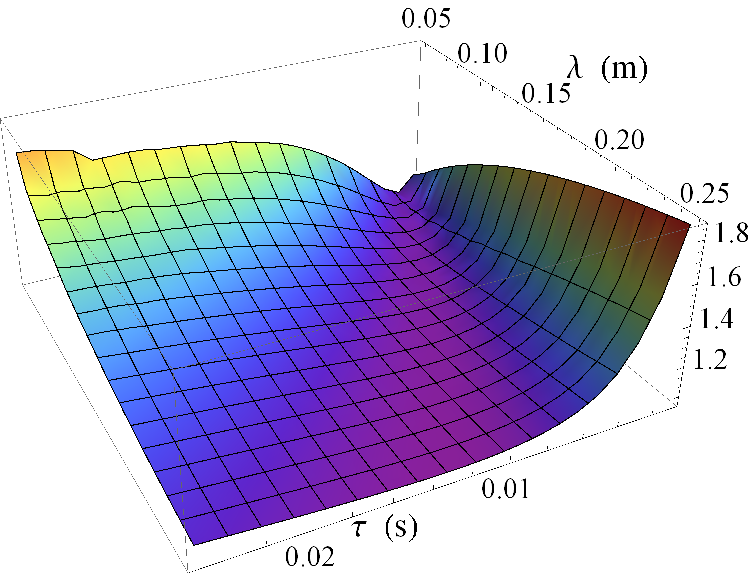}}\hfill
\subfloat[\label{fig10b}]{\includegraphics[height=5.2cm]{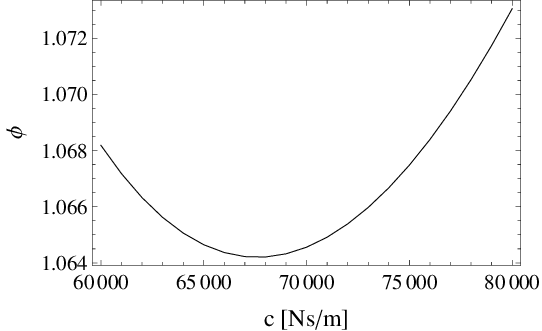}}\caption{The
optimization map (a) of the non-linear system, and the optimization curve (b)
of the linear one. In the optimization process, we set $t_{1}=15$ s$\,$, and
$t_{2}=40$ s.}%
\label{fig10}%
\end{figure}

The results of the optimization process are shown in figures \ref{fig10} for
the nonlinear system (\ref{fig10a}), and, for comparison, for the linear one
(\ref{fig10b}). Interestingly, the nonlinear system results show an optimum
flat valley, allowing to achieve significantly optimized results with several
sets of the design parameters $\tau$ and $\lambda$, providing enhanced
compliance to the different design requirements (i.e. geometrical or material restrictions).

\begin{table}[ptb]
\centering%
\begin{tabular}
[c]{c|c|c|}\cline{2-3}
& Non-Linear & Linear\\\hline
\multicolumn{1}{|c|}{\ Parameters \ } & $%
\begin{array}
[c]{c}%
\tau=0.013\text{ s}\\
\lambda=0.25\text{ m}%
\end{array}
$ & $c=68000$ $\frac{\text{Ns}}{\text{m}}$\\\hline
\multicolumn{1}{|c|}{$F_{M}^{i}$ \ \ \ (N)} & $80255$ & $85271$\\
\multicolumn{1}{|c|}{$F_{rms}^{i}$\ \ (N)} & $16313$ & $17342$\\\hline
\end{tabular}
\caption{Comparison between nonlinear and linear optimized systems results.}%
\label{tab2}%
\end{table}

\begin{figure}[ptbh]
\centering\subfloat[\label{fig11a}]{\includegraphics[height=5.2cm]{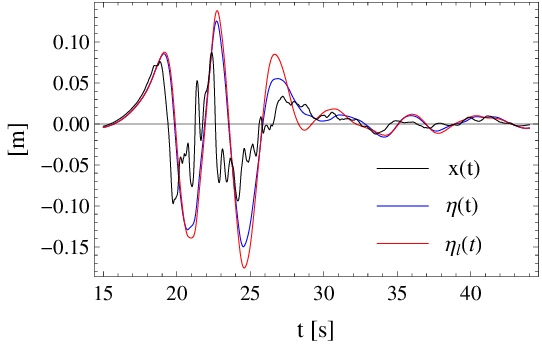}}\hfill
\subfloat[\label{fig11b}]{\includegraphics[height=5.2cm]{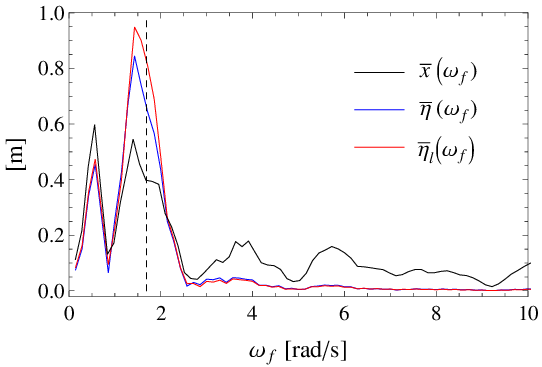}}\caption{The
time history (a) and spectral analysis (b) of: the Central Italy earthquake
2016 (black curves); the inertial mass dispacement $\eta$ for the nonlinear
(blue curves) and linear systems (red curves). The dashed line in (b)
represent the system first natural frequency $\omega_{1}$.}%
\label{fig11}%
\end{figure}

A numerical comparison in terms of optimization results is found in Table
\ref{tab2}. Interestingly, the nonlinear behavior of the RLRB device allows to
reduce both $F_{\max}^{i}$ and $F_{rms}^{i}$ of about $6.3$\%, compared to the
linear system. Such a result is also shown in figures \ref{fig11} where the
displacement time history and spectral analysis is shown for both the systems,
compared to the earthquake data. The smoother behavior of the nonlinear system
shown in fig. \ref{fig11a} (blue line) compared to the linear system (red
line) clearly entails lower inertial effects, and in turn lower stresses for
the structural elements. Furthermore, the system equipped with the nonlinear
RLRB suffers smaller (about $8.4$\%) maximum\ relative base displacement
$\left\vert z\right\vert _{\max}$, compared to the linear isolator, thus
reducing the risk of lateral impact with other structures. Similarly, from
fig. \ref{fig11b}, we observe that, although both the systems are able to
filter the high frequency spectrum of the seismic event, the nonlinear device
(blue curve) still behaves better than the linear one (red curve) even close
to $\omega_{1}$.

\section{Conclusions}

We investigated the dynamic behavior of RLRB seismic isolators, in which the
viscoelastic rolling friction between the rigid cylinders and the rubber
layers leads to a nonlinear damping. We found that the viscoelastic damping
force is a bell-shaped function of the relative velocity of the moving parts
(the ground and the building). Specifically, the damping force increases with
increasing relative velocity of the moving parts up to a peak value of the
damping force. At larger relative velocities, the damping force decreases.
Such a strongly nonlinear trend is controlled by the viscoelastic material
relaxation time, and the rigid cylinders spacing, which indeed dramatically
affect the overall system behavior. Specifically, depending on whether the
RLRB operating condition lies on the increasing or decreasing portion of the
damping curve, strongly nonlinear aperiodic behavior can be observed. We
investigate both the effect of the excitation frequency, as well as the
specific set of parameters.

A real seismic event has been numerically reproduces in order to test the
model, based on the Central Italy earthquake of October 2016. Indeed, an
optimization procedure has been performed to minimize a performance index
taking into account both the maximum instantaneous value and the root mean
square value of the inertial loads history. Similarly, the behavior of an
equivalent linearized system is investigated for comparison. Results show that
the nonlinear system is able to sufficiently reduce both the instantaneous and
averaged inertial load value with respect to the linear system, opening the
path to further deeper investigation on similar devices. Indeed, different
sources of nonlinearity in RLRB devices will be further investigated.

\end{document}